\newcommand{\be}{\begin{equation}}\newcommand{\ee}{\end{equation}}
\newcommand{\bea}{\begin{eqnarray}}\newcommand{\eea}{\end{eqnarray}}
\newcommand{\nn}{\nonumber}\newcommand{\p}[1]{(\ref{#1})}
\newcommand{\un}{\underline}
\documentstyle[12pt]{article}
\begin{document}

     \thispagestyle{empty}
\begin{flushright}
JHU-TIPAC-93007\\  BONN-HE-93-05
\end{flushright}

\begin{center} {\bf
\Large{\bf A twistor formulation of the non-heterotic
superstring  with manifest worldsheet supersymmetry}}
\end{center} \vskip 1.0truecm

\centerline{{\bf A. Galperin}$\, ^{\dag}$}
\vskip5mm
\centerline{\it Department of Physics and Astronomy }
\centerline{\it The Johns Hopkins University}
\centerline{\it Baltimore, Maryland 21218, USA}
\vskip5mm
\centerline{\bf and}
\vskip5mm
\centerline{{\bf E.Sokatchev}$\, ^{\ddag}$}
\vskip5mm
\centerline{\it Physikalisches Institut}
\centerline{\it Universit\"at Bonn}
\centerline{\it Nussallee 12, D-5300 Bonn 1, Germany}
 \vskip 1.4truecm \bigskip   \nopagebreak

\begin{abstract}
We propose a new formulation of the $D=3$ type II
superstring which is manifestly invariant under both  target-space
$N=2$ supersymmetry and worldsheet $N=(1,1)$ super
reparametrizations. This gives rise to a set of twistor (commuting
spinor)
variables, which provide a solution to the two Virasoro constraints.
The
worldsheet supergravity fields are shown to play the r\^ole of
auxiliary
fields.
\end{abstract}
 \bigskip \nopagebreak \begin{flushleft} \rule{2
in}{0.03cm} \\ {\footnotesize \ ${}^{\dag}$  On leave from the
Laboratory of Theoretical Physics, Joint Institute for Nuclear
Research, Dubna, Russia}
\\ {\footnotesize \ ${}^{\ddag}$  On leave from the
Institute for Nuclear
Research and Nuclear Energy, Sofia, Bulgaria}
\vskip8mm
 March 1993
 \end{flushleft}

\newpage\setcounter{page}1

\section{Introduction} \label{intr}

During the last few years a new formulation of the
superparticle and the heterotic superstring with a $D=3,4,6,10$
target
space has been developed \cite{STV}-\cite{DGHS}. It has $N=1$ target
supersymmetry
and at the same time manifest worldline (or worldsheet) $N=D-2$ local
supersymmetry. The latter replaces the well-known kappa-symmetry
\cite{S}, \cite{GSW} of
the superparticle (string). Thus kappa-symmetry finds its natural
explanation as an on-shell version of the off-shell local
supersymmetry of the worldsheet.

The key to such formulations is the use of commuting spinor
(``twistor")
variables, as proposed in the pioneering work of Sorokin, Tkach,
Volkov and
Zheltukhin \cite{STV}. These variables emerge in a natural way as the
worldsheet supersymmetry superpatners of the target superspace
Grassmann
coordinates. In this context one obtains a twistor-like solution for
the
null momentum of the massless superparticle (or for one of the
Virasoro vectors of the heterotic superstring) as a bilinear
combination of
the twistors. Thus, the twistor variables turn out to parametrize the
sphere $S^{D-2}$ associated with the above null vector. All of this
is
achieved as a consequence of one of the equations of motion of the
twistor-like superparticle (string), the so-called geometro-dynamical
constraint. It specifies the way the worldsheet superspace ${\cal M}$
is
embedded in the target superspace $\un{\cal M}$. Namely, one requires
that
the odd part of the tangent space to ${\cal M}$ lies entirely within
the
odd part of the tangent space to $\un{\cal M}$ at any point of ${\cal
M}$.
The conditions for this particular embedding are generated
dynamically from
a Lagrange multiplier term in the action.

It is very natural to try to extend the above results to the
non-heterotic
superstring. This means to solve {\it both} Virasoro constraints in
terms
of twistor variables and to interpret the kappa-symmetry of the
theory as
non-heterotic $N=(D-2,D-2)$ worldsheet supersymmetry. However,
changing from one
dimension (the case of the superparticle) or essentially one
dimension (the
case of the heterotic superstring) to two dimensions of the
worldsheet is
far from trivial. An attempt in this direction has
recently been made by Chikalov and Pashnev \cite{CP}. There only the
first half
of this program was achieved. Considering an $N=2$ target superspace,
but
still only $N=(1,0)$ worldsheet supersymmetry, Chikalov and Pashnev
obtained two twistor variables and solved both Virasoro constraints.
At
the same time, their worldsheet possessed only one supersymmetry,
which
could not explain the full kappa-symmetry of the theory and in
addition
broke two-dimensional Lorentz invariance. An interesting feature of
their
formulation was the absence of any worldsheet supergravity fields.

In the present paper we shall make a step further towards the
realization
of the full twistor program. We present a twistor formulation of the
$D=3$
type II (i.e. with $N=2$ target space supersymmetry) non-heterotic
superstring. On the worldsheet we have  $N=(1,1)$ local supersymmetry
and
thus are able to completely eliminate kappa-symmetry. At first sight
our
construction  closely resembles the one in the heterotic case
\cite{DGHS}.
However, significant differences appear in the analysis of the
twistor
constraints, which follow from the geometro-dynamical embedding of
${\cal
M}$ in $\un{\cal M}$. If in the heterotic case it was relatively easy
to
show that as a result of these constraints the twistor variables
parametrized the space $S^{D-2}$, here a carefull study is needed.
The
solution to the twistor constraints now consists of two sectors, a
regular
one, which corresponds to non-trivial superstring motion and a
singular
one, in which the superstring collapses into a superparticle. Another
big
difference is that one needs the full set of worldsheet supergravity
fields
in order to make the superfield action super reparametrization
invariant.
\footnote{In the case of the superparticle no worldline supergravity
fields
appear in the action \cite{GS}. For the heterotic superstring only
one
component of the two-dimensional metric is needed to generate the
second
Virasoro constraint \cite{DIS}, \cite{DGHS}.} However, at the level
of components one
discovers that the worldsheet gravitino is in fact an {\it auxiliary
field}
and can be eliminated via its algebraic field equation. The
worldsheet
metric and the twistor variables compete for the r\^ole of
reparametrization gauge fields. Algebraic elimination of the twistor
variables leads to the familiar Green-Schwarz action. Finally, in the
heterotic case the formalism worked equally well in all the cases
$D=3,4,6,10$. However, in the non-heterotic case the attempt to go
beyond
$D=3$ (thus having extended worldsheet supersymmetry)
causes a problem, namely, the geometro-dynamical constraint starts
producing equations of motion. Understanding this crucial point will
probably give us some new non-trivial insight into the geometric
nature of
the superstring. It will also help us achieve a complete twistor
formulation in all the dimesnions where the classical superstring
exists.

The paper is organized as follows. In section \ref{geom} we explain
the
notation and introduce the basic geometric objects of the worldsheet
and
target superspaces. In section \ref{part} we present the twistor
formulation of the $D=3$ superparticle with $N=2$ target-space and
$N=(1,1)$
worldsheet supersymmetry. This is a simplified version of the
superstring
theory, which helps illustrate some of the new features encountered
here.
In section \ref{dyn} the two terms of the superstring action, the
geometro-dynamical and the Wess-Zumino ones are given and it is shown
how the
former allows one to establish the consistency of the latter. In
section
\ref{comp} we study in detail the component structure of the action.
We
find out which component fields are auxiliary and by eliminating them
arrive at the standard Green-Schwarz action. This analysis crucially
depends on which solution of the twistor constraints we use, the
regular
or the singular one. In the latter case we observe the
string shrinking to a particle.
In the Appendix we find the general solution to the
algebraic twistor constraints, which consists of a regular and a
singular
sector.

\section{Two- and three-dimensional supergeometry}\label{geom}

In this section we shall introduce some basic concepts concerning
$N=2$
superspaces in two and three dimensions. These superspaces will serve
as
the worldsheet and the target space of the superstring,
respectively.

The worldsheet of the type II $D=3$ superstring is a
$2|2$-dimensional
superspace parametrized  by two even and two odd real coordinates
$Z^M
=(\xi^m, \eta^\mu)$, where $m=(0,1)$ and $\mu=(1,2)$. We assume that
it is
endowed by $N=(1,1)$ two-dimensional supergravity. The latter is
described
by a vielbein $E_A{}^M$ and a $SO(1,1)$ Lorentz connection $\omega_A$
which satisfy the following constraint \footnote{We use the following
conventions for the gamma and charge conjugation matrices:
$\gamma^0=C=
\left(\begin{array}{cc}0&1\\1&0\end{array}\right),\;\;
\gamma^1=\left(\begin{array}{cc}0&1\\-1&0\end{array}\right); \;\;
\{\gamma^m,\gamma^n\}=2\eta^{mn}$ with $\eta^{00}=1,
\;\eta^{11}=-1$.}
\be\label{1.1}
\{D_{ \alpha}, D_{ \beta}\} = 2i (\gamma^c C)_{\alpha\beta}
D_c + R_{\alpha\beta}.
\ee
Here $A=(a,\alpha)$ with $a=(0,1)$ and $\alpha=(1,2)$ being $SO(1,1)$
vector
and spinor indices, and the covariant derivatives are
\be\label{1.0}
D_A = E_A{}^M \partial_M + \omega_A.
\ee

Equation \p{1.1} means that we have imposed the following torsion
constraints:
\be\label{1.2}
T_{ \alpha \beta}{}^c = 2i (\gamma^c C)_{\alpha\beta}, \ \
T_{ \alpha \beta}{}^{ \gamma} = 0.
\ee
The complete set of $N=(1,1)$ supergravity constraints and their
consistency
have been studied in \cite{H}.  There it has been shown
that two-dimensional $N=(1,1)$ supergravity can be considered
superfconformally flat (ignoring moduli problems).
This means that there exist superconformal
transformations of the vielbeins and connections which leave \p{1.1}
invariant and which can gauge away all the torsion and curvature
tensors.
For our purposes we shall need the infinitesimal
form of the super-Weyl transformations of the vielbeins:
\bea\label{sweyl}
\delta E_M{}^a &=& \Lambda E_M{}^a,\\
\delta E_M{}^\alpha &=& {1\over 2}\Lambda E_M{}^\alpha - {i\over
2}E_M{}^a
(C^{-1}\gamma_a)^{\alpha\beta} D_\beta \Lambda.\nn
\eea

In what follows we shall often use two-dimensional light-cone
notation. There
one employs $\gamma^{\pm\pm} =
{1\over 2} (\gamma^0 \pm \gamma^1)$ as projection operators for the
two
irreducible
halves of the spinor. Then the light-cone form of \p{1.1} is
\bea
\{D_{+}, D_{+}\} &=& 2i D_{++}  , \nn\\
\{D_{- }, D_{- }\} &=& 2i D_{--}  , \nn\\
\{D_{+}, D_{- }\} &=& R_{+\vert - } \; .\label{lc}
\eea

Our final point about two-dimensional supergravity concerns the
structure
of the covariant derivatives  \p{1.0} taken at the point $\eta =0$.
They
will be used in section \ref{comp}  for evaluating the superstring
component action. From \cite{H}  we learn that in a certain
gauge for the super-Weyl and tangent Lorentz groups one has
\be\label{1.3}
D_{ \alpha}\vert_{\eta=0} = \partial_{ \alpha} + \omega_{ \alpha},
\ \
D_a\vert_{\eta=0} = e_a{}^m(\xi)
\partial_m + \psi_a{}^\mu(\xi) \partial_\mu + \omega_a,
\ee
where $e_a{}^m(\xi)$ and $\psi_a{}^\mu(\xi)$
are the two-dimensional graviton and gravitino fields.
\vskip5mm

Now we pass to the target superspace of the $D=3$ $N=2$ superstring.
It is
a $3|4$-dimensional superspace
parametrized by \footnote{To avoid the proliferation of indices we
use the
same letters to denote similar types of indices on the worldsheet and
in
the target space. The distinction is made by underlining the target
space
ones.} $Z^{\un M} = (X^{\un m}, \Theta^{\un {\mu}}, \bar\Theta^{\un
{\mu}})$, where $\un m =
(0,1,2)$ and $\un\mu = (1,2)$. Note that the Grassmann variables
are combined here into a {\it complex} doublet $\Theta^{\un {\mu}}$
and
its conjugate $\bar\Theta^{\un{\mu}}$ (hence $N=2$).
Our formulation of the superstring will be of a sigma model
type. \footnote{The sigma model nature of the superstring was
revealed in
\cite{HM}.} In such a context one treats the target
space coordinates as worldsheet superfields $Z^{\un M}(Z^M)$. Then
one can
define the following differential one-forms
\be\label{1form1}
E^{\un A} =  dZ^{\un M} E_{\un M}{}^{\un A} (Z).
\ee
Here $E_{\un M}{}^{\un A} (Z)$ are the vielbeins of the target
supergeometry. The flat $D=3$ $N=2$ superspace is characterized by
the
one-forms \footnote{Here $\gamma^{\un a}$ are the ordinary $D=3$
gamma matrices
times the charge conjugation matrix; we use the representaion
$\gamma^{\un 0}=
\left(\begin{array}{cc}1&0\\0&1\end{array}\right),\;\;
\gamma^{\un 1}=
\left(\begin{array}{cc}0&1\\1&0\end{array}\right),\;\;
\gamma^{\un 2}= \left(\begin{array}{cc}1&0\\0&-1\end{array}\right)
$.}
\be\label{flat1form}
E^{\un a} = d X^{\un a} - id\Theta\gamma^{\un a}
\bar\Theta  - id\bar\Theta\gamma^{\un a}
\Theta, \ \ \ E^{\un \alpha} = d \Theta^{\un\alpha}.
\ee
These forms are invariant under target space $D=3$ $N=2$
supersymmetry and
with respect to the worldsheet local symmetries. The
pull-backs of these forms onto the worldsheet are
\be\label{1form}
E_A{}^{\un A} = D_A Z^{\un M} E_{\un M}{}^{\un A} (Z).
\ee

Acting on \p{1form} with the covariant derivative $D_B$ and
performing graded
antisymmetrization in $A,B$ we obtain an important relation which
involves the worldsheet and target superspace torsions:
\be\label{torsion}
D_AE_B{}^{\un C} - (-)^{AB}D_BE_A{}^{\un C} = T_{AB}{}^CE_C{}^{\un C}
-(-)^{A(B+\un B)}E_B{}^{\un B}E_A{}^{\un A}T_{\un{AB}}{}^{\un C}.
\ee
The explicit form of the flat target superspace torsion is
\be\label{2.2}
T_{\un{\alpha\bar\beta}}{}^{\un c} = T_{\un{\bar\alpha\beta}}{}^{\un
c}
= 2i (\gamma^{\un c})_{\un{\alpha\beta}} ,\ \ \ {\rm the\ rest\ =
\ 0.}
\ee

A characteristic feature of the superstring considered as a sigma
model is
the presence of a Wess-Zumino term in the action. It is based on
another
target superspace geometric object, the super two-form
$B_{\un{MN}}(Z^{\un K})$. In the flat case it is given by
\bea
&&B_{\un{\mu n}} = {i\over 2} (\gamma_{\un n}\Theta)_{\un\mu}, \ \
B_{\un{\bar\mu n}} = {i\over 2} (\gamma_{\un n}\bar\Theta)_{\un\mu},
\nn\\
&&B_{\un{\mu\nu}} = -{1\over 2}(\gamma^{\un
n}\Theta)_{\un\mu}(\gamma_{\un
n}\Theta)_{\un\nu}, \ \ B_{\un{\bar\mu\bar\nu}} = -{1\over
2}(\gamma^{\un
n}\bar\Theta)_{\un\mu}(\gamma_{\un n}\bar\Theta)_{\un\nu},
\label{2form}\\
&&{\rm the\ rest\ = \ 0.} \nn
\eea
Its field strength is a three-form,
\be\label{3for}
H_{\un{MNK}} = \partial_{[\un M} B_{\un{NK})},
\ee
where $[\un{MNK})$ means graded antisymmetrization. Using the $D=3$
gamma
matrix identity
\be\label{gammaid}
(\gamma^{\un m})_{(\un{\nu\lambda}}(\gamma_{\un
m})_{\un\gamma)\un\kappa}=0,
\ee
one can show that
\be\label{3f}
H_{\un{m\nu\lambda}} =H_{\un{m\bar\nu\bar\lambda}} = i(\gamma_{\un
m})_{\un{\nu\lambda}}, \ \ \ {\rm the\ rest\ = \ 0.}
\ee
In what follows we shall also need the expression for the three-form
with
tangent space indices
\be\label{tang}
H_{\un{ABC}} = (-)^{(\un B+\un N)\un A + (\un C+\un K)(\un A+\un B)}
E_{\un C}{}^{\un K} E_{\un B}{}^{\un N} E_{\un A}{}^{\un M}
H_{\un{MNK}}.
\ee
Its projections are similar to those in \p{3f}:
\be\label{3.7}
H_{\un{a\beta\gamma}} =H_{\un{a\bar\beta\bar\gamma}} = i(\gamma_{\un
a})_{\un{\beta\gamma}}, \ \ \ {\rm the\ rest\ = \ 0.}
\ee

 Note that the part of the target space geometry involving only the
 pull-backs
(one-forms) and the torsion respects the automorphism group
$U(1)$ of $D=3$ $N=2$ supersymmetry. The latter rotates $\Theta$ by a
phase
factor. A peculiarity of the type II superstring is that its
Wess-Zumino
term
violates this $U(1)$ symmetry, as can be seen from \p{2form}. As
shown in
\cite{CNM}, this is the only way to have a closed three-form in a
type II
superspace. An interesting geometric interpretation of this fact has
been
given in \cite{II}.

\section{The $D=3$ $N=2$ superparticle}\label{part}

In this section we shall present a twistor formulation of the
superparticle moving in $D=3$ $N=2$ superspace. It is the
one-dimensional simplified version of the superstring. It shares some
new features with the superstring and can thus serve as an
introduction to the superstring. Moreover, as we shall see in
subsection \ref{sing}, a specific solution to the superstring twistor
constraints leads to a degenerate form of the superstring, which is
just the superparticle.

  In the traditional
Brink-Schwarz formulation \cite{BS} of the $D=3$ $N=2$ superparticle
one finds two
kappa-symmetries wich gauge away half of the target space Grassmann
coordinates. In a twistor formulation one expects to have two
worldline
local supersymmetries. So, we consider a superworldline parametrized
by
$\tau, \eta^i$, where $i=1,2$ is a doublet index of the $SO(2)$
automorphism of the $N=2$ supersymmetry algebra
\be\{D_i, D_j\}=2i\delta_{ij}\partial_\tau .
\label{p1}\ee

The action we propose for the $D=3$ $N=2$ superparticle is given by
\be
S=\int d\tau d^2\eta P_{i\un a}(D_i X^{\un a}
 - i D_i\Theta\gamma^{\un a}\bar\Theta  -
i D_i\bar\Theta\gamma^{\un a}\Theta ).
\label{p2}
\ee
It  contains the pull-back $E_i{}^{\un a}$ of the invariant one form
of target space supersymmetry (cf. \p{flat1form})
and a Lagrange multiplier.  All superfields in
\p{p2} are unconstrained. As a kinematical restriction we require
that
 the  pull-back $E_i{}^{\un\alpha}$ defining the commuting spinor
 (twistor) variables be a non-vanishing matrix,
\be
D_i\Theta^{\un\alpha}\neq 0.\label{non}
\ee
We  note that this action is invariant under the $N=2$ superconformal
group. \footnote{For a discussion of the extended one-dimensional
superconformal groups see \cite{GS}.}  Our aim
is to study the component content of the above action and to show its
equivalence to the Brink-Schwarz superparticle action \cite{BS}.
Integrating over the worldline Grassmann coordinates we get
\bea
 S&=&\int d\tau [D_1D_2P_{i\un a}(D_iX^{\un a}
 - i D_i\Theta\gamma^{\un a}\bar\Theta
-i D_i\bar\Theta\gamma^{\un a}\Theta) \nn\\
&&+iD_2P_{1\un a}(\partial_\tau X^{\un a}
- i \partial_\tau\Theta\gamma^{\un a}\bar\Theta
- i \partial_\tau\bar\Theta\gamma^{\un a}\Theta
-2D_1\Theta\gamma^{\un a}D_1\bar\Theta) \nn\\
&&-iD_1P_{2\un a}(\partial_\tau X^{\un a}
- i \partial_\tau\Theta\gamma^{\un a}\bar\Theta
- i \partial_\tau\bar\Theta\gamma^{\un a}\Theta
-2D_2\Theta\gamma^{\un a}D_2\bar\Theta)\label{p3}\\
&&+D_1P_{1\un a}(D_1D_2X^{\un a}-iD_1D_2\Theta\gamma^{\un
a}\bar\Theta
-iD_1D_2\bar\Theta\gamma^{\un a}\Theta
+iD_1\Theta\gamma^{\un a}D_2\bar\Theta
+iD_1\bar\Theta\gamma^{\un a}D_2\Theta)\nn\\
&&+D_2P_{2\un a}(D_1D_2X^{\un a}-iD_1D_2\Theta\gamma^{\un
a}\bar\Theta
-iD_1D_2\bar\Theta\gamma^{\un a}\Theta
-iD_1\Theta\gamma^{\un a}D_2\bar\Theta
-iD_1\bar\Theta\gamma^{\un a}D_2\Theta)\nn\\
&&-iP_{1\un a}({1\over 2}\partial_\tau D_2X^{\un a}
-i\partial_\tau D_2\Theta\gamma^{\un a}\bar\Theta
+i\partial_\tau\Theta\gamma^{\un a}D_2\bar\Theta
+2D_1D_2\Theta\gamma^{\un
a}D_1\bar\Theta + h.c.)\nn\\
&&+iP_{2\un a}({1\over 2}\partial_\tau D_1X^{\un a}
-i\partial_\tau D_1\Theta\gamma^{\un a}\bar\Theta
+i\partial_\tau\Theta\gamma^{\un a}D_2\bar\Theta
-2D_1D_2\Theta\gamma^{\un
a}D_2\bar\Theta + h.c.)]_{\eta=0} . \nn
\eea

The variation with respect to the component $D_1D_2P_{i\un a}$
produces
the auxiliary field equations (we omit the subscript $\eta=0$)
\be
D_iX^{\un a} - i D_i\Theta\gamma^{\un a}\bar\Theta
- i D_i\bar\Theta\gamma^{\un a}\Theta=0 .
\label{p4}\ee
The variation with respect to the components $D_1P_{1\un a}$ and
$D_2P_{2\un a}$
defines the auxiliary component $D_1D_2 X^{\un a}$ and also leads to
one of
the twistor constraints
\be
D_1\Theta\gamma^{\un a}D_2\bar\Theta +
D_1\bar\Theta\gamma^{\un a}D_2\Theta=0 .\label{p5}
\ee
Further, varying with respect to the sum $D_2P_{1\un a}+D_1P_{2\un
a}$ we
get the other twistor constraint
\be
D_1\Theta\gamma^{\un a}D_1\bar\Theta - D_2\Theta\gamma^{\un
a}D_2\bar\Theta=0 .
\label{p6}\ee
The difference $i(D_2P_{1\un a}-D_1P_{2\un a})$ is identified with
the
particle's momentum $p_{\un a}$.

Finally, the last two terms in the component action \p{p3} can be
simplified by using the auxiliary field equations \p{p4} and the
resulting
action  takes the form
\bea
S&=&\int d\tau [p_{\un a}(\partial_\tau X^a
-i\partial_\tau\Theta\gamma^{\un a}\bar\Theta
-i\partial_\tau\bar\Theta\gamma^{\un a}\Theta
-D_i\Theta\gamma^{\un a}D_i\bar\Theta) \nn\\
&&+2P_{1\un a}(D_2\Theta\gamma^{\un a}\partial_\tau\bar\Theta+
D_2\bar\Theta\gamma^{\un a}\partial_\tau\Theta
-iD_1\Theta\gamma^{\un a}D_1D_2\bar\Theta
-iD_1\bar\Theta\gamma^{\un a}D_1D_2\Theta )\nn\\
&&-2P_{2\un a}(D_1\Theta\gamma^{\un a}\partial_\tau\bar\Theta
+D_1\bar\Theta\gamma^{\un a}\partial_\tau\Theta
+iD_2\Theta\gamma^{\un a}D_1D_2\bar\Theta
+iD_2\bar\Theta\gamma^{\un a}D_1D_2\Theta )] . \label{p7}\eea

Here the twistor variables $D_i\Theta^{\un \alpha }$
and $D_i\bar\Theta^{\un \alpha }$ are restricted by the
constraints \p{p5} and \p{p6}. Our next step is to solve these
constraints.
Using the explicit representation for the $D=3$ gamma matrices in the
light-cone basis
\be
\gamma^{++} = \left(\matrix{1 & 0 \cr 0 & 0 \cr} \right),
\ \gamma^{--} =
\left(\matrix{0 & 0 \cr 0 & 1 \cr} \right), \ \gamma^{+-} =
\left(\matrix{0 & 1 \cr 1 & 0 \cr} \right) \label{A3}
\ee
we easily find the general solution  to the twistor
constraints \p{p5} and \p{p6} under the assumption \p{non}:
\be
D_1\Theta^{\un\alpha}=\lambda^{\un \alpha}, \;\;\;
D_2\Theta^{\un\alpha}=i s\lambda^{\un \alpha}; \;\; s=\pm 1,
\label{p12}\ee
where $\lambda^{\un \alpha}$ is an arbitrary {\it complex
nonvanishing}
spinor.

Substituting this solution into the action \p{p7} we get
\be
S=\int d\tau [p_{\un a}(\partial_\tau X^a
-i\partial_\tau\Theta\gamma^{\un a}\bar\Theta
-i\partial_\tau\bar\Theta\gamma^{\un a}\Theta
-2\lambda\gamma^{\un a}\bar\lambda)
 -2P_{\un a}
\bar\chi\gamma^{\un a}\lambda -2\bar P_{\un a}
\chi\gamma^{\un a}\bar\lambda ],
\label{p13}\ee
where
$$
P_{\un a} = P_{2\un a}-is P_{1\un a}, \ \
\chi^{\un \alpha}\equiv \partial_\tau\Theta^{\un \alpha}
+sD_1D_2\Theta^{\un\alpha} .
$$

Now we remark that the tri-linear term in this action is purely
auxiliary. Indeed, the variation with respect to $P_{\un a}$
leads to the equation
\be
\bar\chi\gamma^{\un a}\lambda=0 .
\label{p14}\ee
Since the commuting spinor $\lambda$ is non-vanishing, the only
solution to
this equation is $\bar\chi=0$. The variation with respect to
$\bar\chi$ in
\p{p13} implies
\be
P_{\un a}(\gamma^{\un a}\lambda)_{\un\alpha}=0 .
\ee
The general solution of this twistor equation is
\be
P_{\un a}=\rho(\tau)\lambda\gamma^{\un a}\lambda \label{rho}
\ee
with an arbitrary complex
function $\rho(\tau)$. However, the right hand side of
\p{rho} is an obvious gauge invariance of the action \p{p13} due to
the
gamma matrix identity \p{gammaid}.
Hence the two last terms in \p{p13} vanish.

Finally, we vary with respect to the twistor variable $\lambda$:
\be
p_{\un a}(\gamma^{\un a}\bar\lambda)_{\un\alpha}=0\  \Rightarrow
\ p^{\un
a}=\bar\mu(\tau)\bar\lambda\gamma^{\un a}\bar\lambda .
\ee
The fact that  the  particle momentum is real and non-vanishing then
implies
\be
\bar\mu(\tau)\bar\lambda\gamma^{\un a}\bar\lambda=
\mu(\tau)\lambda\gamma^{\un a}\lambda .
\ee
The solution to this equation is
\be \bar\lambda^{\un \alpha}=e^{i\phi}\lambda^{\un \alpha},
\;\; \bar\mu=e^{-2i\phi}\mu .
\ee
It implies that on shell the complex spinor $\lambda^{\un\alpha}$
becomes {\it real}
modulo a phase. The arbitrary phase $\phi$ corresponds to
the $SO(2)$ subgroup of the superconformal invariance of the action
\p{p2}
and can be completely gauged away. Then we can replace the twistor
combination $\lambda\gamma^{\un a}\bar\lambda$ in \p{p13} by
$|\mu|^{-1}
p^{\un a}$ and obtain the standard Brink-Schwarz superparticle
action.

The conclusion is that the action \p{p2} is equivalent to the
Brink-Schwarz action upon eliminating the auxiliary fileds (including
the
twistor variables) and fixing certain gauges. An unusual feature
compared to the twistor superparticle of \cite{STV} is the presence
of two twistors (the real and imaginary parts of $\lambda$) instead
of only one. As we shall see in subsection \ref{sing}, it is this
form of the superparticle which appears as a degenerate case of the
non-heterotic superstring.

\section{The $D=3$ $N=2$ superstring action}\label{dyn}

The twistor superstring action consists of two terms
\be\label{action}
S = S_{GD} + S_{WZ}.
\ee
The first one resembles very much the superparticle action of section
\ref{part}:
\be\label{4.1}
S_{GD} = \int d^2\xi d^{2}\eta P_{\un a}{}^{ \alpha}
E_{ \alpha}{}^{\un a} .
\ee
The variation with respect to the Lagrange
multiplier $P_{\un a}{}^{ \alpha}$ leads to the following
geometro-dynamical \footnote{``Geometro-" because it determines the
way
the worldsheet is embedded as a surface in the target superspace;
``dynamical" because it is derived from the superstring
action.} constraint on
the target space coordinates treated as worldsheet
superfields $Z^{\un M}(Z^M)$:
\be\label{2.1}
E_{  \alpha}{}^{\un a}  = 0.
\ee

The meaning of eq. \p{2.1} is that the pull-back of the target
superspace
vector vielbein onto the spinor directions of the worldsheet must
vanish.
In other words, if one considers the worldsheet as a
$2|2$-dimensional
hypersurface embedded in the $3|4$ target superspace, then there
should be no projections of the target space even directions onto the
worldsheet odd ones. Using \p{torsion}, \p{1.2} and \p{2.2},
we obtain the important consequence
\be\label{2.3}
2( \gamma^cC)_{\alpha\beta} E_c{}^{\un a} =
E_{ \alpha}\gamma^{\un a} \bar E_{ \beta} + \bar E_{
\alpha}\gamma^{\un a}
E_{ \beta}.
\ee
In two-dimensional light-cone notation \p{2.3} reads
\bea\label{m1}
&&E_+\gamma^{\un a} \bar E_+ = E_{++}{}^{\un a}, \\
&&E_-\gamma^{\un a} \bar E_- = E_{--}{}^{\un a}, \label{m2}\\
&&E_+\gamma^{\un a} \bar E_- + E_-\gamma^{\un a} \bar E_+ = 0.
\label{m3}
\eea
These equations are constraints on the superfields $Z^{\un M}(Z^M)$.
In
particular, they imply algebraic restrictions on the first components
in
the $\eta$ expansion of the spinor-spinor pull-backs
\be\label{twistors}
{\cal E}_\alpha{}^{\un\alpha} \equiv
E_\alpha{}^{\un\alpha}\vert_{\eta=0}.
\ee
These are commuting spinors (with respect to the two- and
three-dimensional
Lorentz groups), which we shall call
``twistor variables". In  section \ref{comp} we shall show that as a
result of
these restrictions the first components in the $\eta$ expansion of
the vectors
$E_{\pm\pm}{}^{\un a}$ defined by \p{m1} and \p{m2} are lightlike,
\be\label{pullb}
{\cal E}_{\pm\pm}{}^{\un a} = E_{\pm\pm}{}^{\un a}\vert_{\eta = 0}:
\ \ \
({\cal E}_{++}{}^{\un a})^2 = ({\cal E}_{--}{}^{\un a})^2 = 0.
\ee
In other words, one of the main purposes of the geometro-dynamical
constraint
\p{2.1} is to provide a solution to the Virasoro constraints of the
superstring
in terms of the twistor variables \p{twistors}.

The presence of the geometro-dynamical term \p{4.1} makes it possible
to
introduce the leading term in the superstring
action in the form of a generalized Wess-Zumino term. The latter
requires
certain consistency conditions, and we shall show that they are
satisfied
as a consequence of the geometro-dynamical constraint \p{2.1}.

The Wess-Zumino term has the following form
\be\label{3.1}
S_{WZ} = \int d^2\xi d^{2}\eta \;(-)^{MN+M+N} P^{MN}(B_{MN} -
E_M{}^{a}E_N{}^{b}
\epsilon_{ab} A -\partial_M Q_N).
\ee
Here $P^{MN}$ is a graded-antisymmetric Lagrange multiplier, $Q_M$ is
another Lagrange multiplier, $B_{MN} =
(-)^{(N+\un N)M}E_N{}^{\underline N}E_M{}^{\underline M}
B_{\underline {MN}}$ is
the pull-back of the target superspace two-form, $E_{M}{}^{a}$ are
worldsheet vielbeins. The quantity $A$ is to be found from the
consistency
conditions below. Varying with respect to $P^{MN}$ leads to the
equation
of motion
\be\label{3.2}
B_{MN} - E_{M}{}^{a}E_{N}{}^{b} \epsilon_{ab} A = \partial_{[M}
Q_{N)}.
\ee
The meaning of this equation is that the pull-back of the two-form
becomes
almost ``pure gauge" on shell. The consistency
condition following from \p{3.2} is that the graded curl of the
left-hand side of eq.\p{3.2} must vanish. Thus we obtain
\be\label{3.3}
H_{MNK} = \partial_{[K} B_{MN)} =
2\partial_{[K}E_M{}^a E_{N)}{}^b\epsilon_{ab} A + (-)^{K(M+N)}
E_{[M}{}^a E_{N}{}^b
\epsilon_{ab} \partial_{K)} A,
\ee
where $H_{MNK}$ is the pull-back of the three-form. It is convenient
to pass to tangent space indices in eq. \p{3.3}, i.e. to multiply it
by
two-dimensional vielbeins. Using the expression for the worldsheet
torsion
\be\label{wstor}
T_{AB}{}^C = -(-)^{A(B+N)} E_B{}^N E_A{}^M \partial_ME_N{}^C +
\omega_{AB}{}^C
-(-)^{AB} (A\leftrightarrow B)
\ee
one can rewrite \p{3.3} as follows:
\be\label{3.4}
H_{ABC} = (-T_{[BA}{}^d +2\omega_{[BA}{}^d)\delta_{C)}^e
\epsilon_{de} A +
\delta_{[B}^d\delta_{C}^e\epsilon_{de}D_{A)} A \; ,
\ee
where $H_{ABC}$ is the pull-back of the three-form
\be\label{3form}
H_{ABC} = (-)^{(B+\un B)A + (C+\un C)(A+B)} E_C{}^{\un C} E_B{}^{\un
B}
E_A{}^{\un A} H_{\un{ABC}} \; .
\ee

Let us study the different projections of the condition
\p{3.4}.  First we consider the projections of its left-hand side.
 If we take all the
indices spinor and use the geometro-dynamical equation \p{2.1},
\footnote{Here and in what follows we frequently use the constraint
\p{2.1} or its corollary \p{2.3} in the Wess-Zumino action term. This
actually means that we produce terms proportional to the constraint,
which
can be absorbed
into a redefinition of the Lagrange multiplier in \p{4.1}.} we find
\be\label{3.6}
H_{ \alpha \beta \gamma} = E_{ \alpha}{}^{\un{\alpha
}}E_{ \beta}{}^{\un{\beta }}E_{ \gamma}{}^{\un{\gamma }}
H_{\un{\alpha \beta \gamma }} + 3E_{ \alpha}{}^{\un{\alpha
}}E_{ \beta}{}^{\un{\beta }}E_{ \gamma}{}^{\un{\bar\gamma }}
H_{\un{\alpha \beta \bar\gamma }} + {\rm c.c.} = 0
\ee
as a consequence of \p{3.7}. In accordance with this
the right-hand side of eq. \p{3.4} vanishes identically for this
choice of
the indices.

Further, let us take one vector and two spinor indices in
\p{3.4}. Substituting  \p{3.7}  into \p{3form} and using
\p{m1}-\p{m3} and \p{gammaid}, we find
\be\label{3.9}
H_{++\vert +\vert +} =i(E_+ \gamma_{\un a} E_{+}) E_{++}{}^{\un a} +
{\rm
c.c.} =i (E_+ \gamma_{\un a} E_{+})(E_+ \gamma^{\un a} \bar E_{+}) +
{\rm
c.c} = 0;
\ee
\be\label{3.91}
H_{++\vert +\vert -} =i(E_- \gamma_{\un a} E_{+}) E_{++}{}^{\un a} +
{\rm
c.c.} = i(E_- \gamma_{\un a} E_{+})(E_+ \gamma^{\un a} \bar E_{+}) +
{\rm
c.c}
\ee
$$
= -{i\over 2}(E_+ \gamma_{\un a} E_{+})(E_- \gamma^{\un a} \bar
E_{+}) +
{\rm
c.c}= {i\over 2}(E_+ \gamma_{\un a} E_{+})(E_+ \gamma^{\un a} \bar
E_{-})
+ {\rm
c.c} =0;
$$
Similarly, we obtain
\be\label{3.90}
H_{--\vert -\vert -} =H_{--\vert +\vert -} =0.
\ee

The only non-vanishing pull-backs with one vector and two spinor
indices are
\be\label{3.92}
H_{--\vert +\vert +} =i(E_+ \gamma_{\un a} E_{+}) E_{--}{}^{\un a} +
{\rm
c.c.} = i(E_+ \gamma_{\un a} E_{+})(E_- \gamma^{\un a} \bar E_{-}) +
{\rm
c.c}
\ee
$$
= -2i(E_+ \gamma_{\un a} E_{-})(E_+ \gamma^{\un a} \bar E_{-}) + {\rm
c.c};
$$
\be\label{3.93}
H_{++\vert -\vert -} =i(E_- \gamma_{\un a} E_{-}) E_{++}{}^{\un a} +
{\rm
c.c.} = i(E_- \gamma_{\un a} E_{-})(E_+ \gamma^{\un a} \bar E_{+}) +
{\rm
c.c}
\ee
$$
= -2i(E_- \gamma_{\un a} E_{+})(E_- \gamma^{\un a} \bar E_{+}) + {\rm
c.c} = - H_{--\vert +\vert +}
$$
If we compare these expressions for the pull-backs of the three-form
with the
right-hand side of eq. \p{3.4} and use \p{1.2},
we see complete agreement, provided that the quantity $A$ is given by
\be\label{3.15}
A = {1\over 8}  \left(E \gamma_a \gamma_{\un a}
E + {\rm c.c.}\right)\epsilon^{ab} E_{b}{}^{\un a}.
\ee

The remaining possibility in eq. \p{3.4} is to have two vector and
one
spinor index. This does not lead to new relations, since the
component
$H_{ab \gamma}$ of the pull-back of the three-form is determined by
the
Bianchi identity
\be
D_{[a}H_{ \alpha \beta \gamma)} + T_{[a \beta}{}^E
H_{E \gamma \delta)} = 0
\ee
and thus automatically agrees with the right-hand side of \p{3.4}.

This concludes the verification of the consistency of our Wess-Zumino
term.
We have seen that the pull-back of the two-form itself is not closed
($dB
\neq 0$), but this can be corrected by an appropriately chosen term
with
$A$ given in \p{3.15}.

We note that the action term \p{3.1} is invariant
under the superconformal transformations \p{sweyl}. Indeed,
we see that the vielbein factor in front of $A$ in \p{3.1}
transforms as a density of weight $+2$.
The twistor vector $E\gamma^a \gamma_{\un a} E$ and
its conjugate are densities of weight $-1$.
As to the vectors $E_{a}{}^{\un a}$,
their transformation laws are less trivial. Take, for instance,
\be\label{3.17''}
\delta E_{++}{}^{\un a} = -\Lambda E_{++}{}^{\un a} +i D_{+} \Lambda
E_{+}{}^{\un a}.
\ee
The first term in \p{3.17''} provides the
weight $-1$ needed to compensate the other two factors. The second
term is
proportional to $E_\alpha{}^{\un a}$, which vanishes according to
\p{2.1}.
In other words, this second term can be compensated by a suitable
transformation of the Lagrange multiplier $P_{\un a}{}^{ \alpha}$ in
the action term \p{4.1}. As to the term \p{4.1} itself, its super
Weyl
invariance is assured by ascribing a certain weight to the Langrange
multiplier.

Another remark concerns the $U(1)$ automorphism of $D=3$
$N=2$ supersymmetry. The term $S_{GD}$ of our action respects this
symmetry, whereas $S_{WZ}$ does not.

\section{The component action}\label{comp}

In this section we shall obtain the component
expression of the two terms \p{4.1} and \p{3.1} of the superstring
action.
We shall show that the Wess-Zumino term \p{3.1} is reduced to the
usual
superstring action of Green-Schwarz type. The geometro-dynamical term
\p{4.1} will
turn out to be purely auxiliary.

\subsection{The Wess-Zumino term}\label{WZT}

We begin with the Wess-Zumino term \p{3.1}. The variation with
respect to the
Lagrange multiplier $Q_M$ produces the equation
\be\label{3.17}
\partial_N P^{NM} = 0.
\ee
Its general solution is:
\be\label{3.18}
P^{MN} = \partial_K \Sigma^{MNK} +\delta^M_k\delta^N_l \epsilon^{kl}
T
\;\eta^{2}.
\ee
It consists of two parts. The first one has the form of a pure gauge
transformation with graded antisymmetric parameter
$\Sigma^{MNK}$.  We are sure that this is an invariance of the action
because the
consistency condition \p{3.4} holds. \footnote{Note that \p{3.4} is a
consequence of the geometro-dynamical constraint \p{2.1}. This means
that
the $\Sigma$ gauge invariance is achieved by a suitable
transformation of
the Lagrange multiplier in \p{4.1}.} It is easy to see that almost
all of
the components of $P^{MN}$ can be gauged away by a $\Sigma$
transformation
without using any parameters with space-times derivatives. Hence, one
is
allowed to use such a gauge in the action. The only remaining
non-trivial
part of $P^{MN}$ is the second term in the solution \p{3.18}. It
contains
a coefficient $T$, which is restricted by \p{3.17} to be an arbitrary
constant,
\be\label{3.19}
\partial_{++} T =  \partial_{--} T =  0.
\ee

Inserting the solution \p{3.18} back into the action term \p{3.1} and
doing the $\eta$
integral with the help of the Grassmann delta function $\eta^2$
present in
\p{3.18} we obtain
\be\label{3.20}
S_{WZ} = \int d^2\xi \;T\;\epsilon^{mn}\left(B_{mn} - {1\over 2}
E_m{}^{a}E_n{}^{b} \epsilon_{ab}\;A\right)_{\eta=0} .
\ee
To evaluate the various objects at $\eta=0$ we use eq. \p{1.3}. Thus,
$E_m{}^{a}\vert_0 = e_m{}^{a}(\xi)$, so the factor in front of $A$
becomes
simply $\det e$. The quantity $A$ from \p{3.15} takes the form
\be\label{A'}
A\vert_{\eta=0} = {1\over 8}  \left({\cal E} \gamma_a \gamma_{\un a}
{\cal E} + {\rm c.c.}\right)\epsilon^{ab} {\cal E}_{b}{}^{\un a},
\ee
where
\be\label{pb}
{\cal E}_\alpha{}^{\un\alpha} = E_\alpha{}^{\un\alpha}\vert_0,
\ \ {\cal
E}_a{}^{\un a} = E_a{}^{\un a}\vert_0 = \left[(e_a{}^m\partial_m +
\psi_a{}^\mu D_\mu)
Z^{\un M} E_{\un M}{}^{\un a}\right]_0 = \left[e_a{}^m\partial_m
Z^{\un M}
E_{\un M}{}^{\un a}\right]_0 .
\ee
Note that the only place in \p{3.20} where the gravitino field
$\psi_a{}^\mu$ could occur is in \p{pb}, but even there it dropped
out as a
consequence of the geometro-dynamical equation \p{2.1}.

At this point we are going to use the solution to the $\eta=0$ part
of the
constraint \p{2.3}. This is the constraint on the twistor variables
${\cal
E}_\alpha{}^{\un\alpha}$ and appears as a component of
the geometro-dynamical term \p{4.1} (see  subsection \ref{geo-d}). As
explained in the Appendix, the twistor constraint has two solution, a
regular and a singular one. The regular solution is obtained under
the
assumption that the twistor matrix ${\cal
E}_\alpha{}^{\un\alpha}$ is non-degenerate,
\be\label{non-deg}
\det \parallel {\cal
E}_\alpha{}^{\un\alpha}\parallel \neq 0
\ee
and has the form
\be\label{matrsol}
{\cal E}_\alpha{}^{\un\alpha} = e^{i\phi} \left(\begin{array}{c}
\lambda_{+}{}^{\un\alpha} \\ i\lambda_{- }{}^{\un\alpha}
\end{array} \right).
\ee
Here $\phi(\xi)$ is an arbitrary phase. The spinors $\lambda$ in
\p{matrsol} are real,
\be\label{real}
\bar \lambda_{+}{}^{\un\alpha} = \lambda_{+}{}^{\un\alpha}, \ \ \bar
\lambda_{- }{}^{\un\alpha} = \lambda_{- }{}^{\un\alpha}
\ee
and satisfy two further relations:
\be
\lambda_{+} \gamma^{\un a} \lambda_{+} = {\cal E}_{++}{}^{\un
a}, \ \ \
\lambda_{- } \gamma^{\un a} \lambda_{- } = {\cal E}_{--}{}^{\un a}.
\label{twcon}
\ee
These equations give expressions for the vectors ${\cal
E}_{\pm\pm}{}^{\un
a}$ in terms of the twistor variables $E_\alpha{}^{\un\alpha}$. Using
\p{gammaid}, one sees that the vectors in \p{twcon} are lightlike,
\be\label{lightl}
({\cal E}_{++}{}^{\un a})^2 =({\cal E}_{--}{}^{\un a})^2 = 0.
\ee
Thus, we see that the lowest-order component of the twistor
constraint
\p{2.3} has reduced the $2\times 2$ complex twistor matrix to two
independent real twistors $\lambda_\pm{}^{\un\alpha}$, in terms of
which the
superstring Virasoro constraints \p{lightl} are solved.

In this subsection we shall restrict ourselves to the regular
solution
\p{matrsol}. The case of the singular one (which corresponds to
the case of a string collapsed into a particle) will be treated in
subsection \ref{sing}. So, putting the above expressions in
\p{A'} and then in \p{3.20}, we obtain
\be\label{3.21}
S_{WZ} = T\int d^2\xi \left( \epsilon^{mn}B_{mn} -
{1\over 2}\det e\; \cos 2\phi\; {\cal E}_{++}{}^{\un a}
{\cal E}_{--\un a}\right).
\ee
We see that this expression is almost identical with
the usual Green-Schwarz type II superstring action, if the constant
$T$
is interpreted as the string tension. \footnote{The mechanism
where the string tension appears as an integration constant was
proposed in a different context in \cite{}}
The only difference is in the factor
containing the auxiliary scalar field $\phi(\xi)$. In
subsection \ref{geo-d} we shall see that $\phi$ does not appear in
the
geometro-dynamical term \p{4.1} of the
superspace superstring action. This is not surprising, since $\phi$
appears as the parameter of a $U(1)$ transformation, and $S_{GD}$
respects
this symmetry. Therefore we can vary with respect to $\phi$
in \p{3.21} and obtain the following field equation
\be\label{3.24}
\sin 2\phi\;{\cal E}_{++}{}^{\un a}
{\cal E}_{--\un a} = 0.
\ee
Using the twistor expressions \p{twcon}, it is not hard to show that
\be\label{indmet}
{\cal E}_{++}{}^{\un a} {\cal E}_{--\un a} =
(\lambda_{+} \gamma^{\un a}
\lambda_{+})(\lambda_{-} \gamma_{\un a} \lambda_{-}) =
(\det \parallel
\lambda_\alpha{}^{\un\alpha}\parallel)^2 = -e^{-4i\phi}
(\det \parallel {\cal
E}_\alpha{}^{\un\alpha}\parallel)^2 .
\ee
Note that the first term in \p{indmet} is in fact proportional to the
determinant of the induced two-dimensional metric of the
superstring.
Since in this subsection we assume that the twistor matrix is
non-singular
(see \p{non-deg}), we conclude that the solution to \p{3.24} is
\be\label{3.25}
\phi = 0.
\ee
Putting this solution back into the action \p{3.21} we obtain
\be\label{3.26}
S_{WZ} = T\int d^2\xi \left( \epsilon^{mn}B_{mn} - {1\over 2} \det e
\;
{\cal E}_{++}{}^{\un a}
{\cal E}_{--\un a} \right).
\ee
This is the action of a type II $D=3$ Green-Schwarz superstring. In
it the
twistor variables are not present any more, they have been eliminated
through the algebraic relations \p{twcon}.

We have seen that in the process of derivation of the action \p{3.26}
the
two-dimensional gravitino field dropped out (see \p{pb}),
although in the original superfield form \p{3.1} it was needed to
maintain the invariance with respect to the local supersymmetry
transformations on the worldsheet. Despite the absence of the
gravitino, this local supersymmetry is still present in the component
term
\p{3.26}, but now in the non-manifest form of kappa symmetry. In
order to
see how local supersymmetry is transformed into kappa symmetry on
shell
let us consider the supersymmetry variation of the target superspace
coordinates $z^{\un M} = Z^{\un M}\vert_{\eta = 0}$
\be\label{susy}
\delta z^{\un A} \equiv \delta z^{\un M} E_{\un M}{}^{\un A} =
\epsilon^\alpha D_\alpha Z^{\un M} E_{\un M}{}^{\un A}\vert_{\eta =
0} =
\epsilon^\alpha {\cal E}_\alpha{}^{\un A}.
\ee
{}From the geometro-dynamical equation \p{2.1} follows
\be\label{susy'}
\delta z^{\un a} = 0, \ \ \delta z^{\un \alpha} = \epsilon^\alpha
{\cal E}_\alpha{}^{\un \alpha}.
\ee
Let us now introduce an anticommuting parameter $\kappa_{\un
\alpha}{}^a$ carrying a $D=2$ vector and a $D=3$  spinor index by
substituting
\be\label{kappa}
\epsilon_\alpha =  (\gamma_a)_\alpha{}^\beta {\cal
E}_\beta{}^{\un\beta}\kappa_{\un\beta}{}^a .
\ee
Putting this in \p{susy'}, using the solution to the constraints on
the
twistor matrix, eq. \p{3.25} and the Fierz identity for the
three-dimensional gamma matrices, we obtain
\be\label{k}
\delta z^{\un \alpha} = (\gamma_{\un a})^{\un{\alpha\beta}} ({\cal
E}_{++}{}^{\un a}\kappa_{\un\beta}{}^{++}  -{\cal E}_{--}{}^{\un
a}\kappa_{\un\beta}{}^{--}).
\ee
Equation \p{k}  coincides with  the kappa symmetry
transformations of the type II superstring \cite{GSW}.

Note an interesting feature of the transition from the action term
\p{3.21} to the final form \p{3.26}. The former is not invariant
under the
$U(1)$ automorphism of $D=3$ $N=2$ supersymmetry. In the first term
in
\p{3.21} this is due to the $U(1)$ non-invariant two-form. In the
second term in \p{3.21} the only object which breaks $U(1)$ is the
phase $\phi$.
Indeed, looking at \p{matrsol}, one sees that $\phi$ is shifted by
the
$U(1)$ transformation of the index $\un\alpha$. However, once this
field
has been eliminated from the action, one obtains the peculiar mixture
of a
non-invariant and an invariant term in \p{3.26}, characteristic for
the
type II superstring.

\subsection{The geometro-dynamical term}\label{geo-d}

In the previous subsection we saw that the usual superstring action
is
essentially contained in the Wess-Zumino term. Here we shall show
that the
r\^ole of the geometro-dynamical term \p{4.1} is purely auxiliary,
i.e.
that it
only leads to algebraic constraints on the component fields. Among
them
are the twistor constraints reducing the twistor matrix to
the two light-like vectors from the Virasoro constraints. Another of
these
constraints will allow us to express the two-dimensional gravitino
field in
terms of the
derivatives of the Grassmann coordinates $\theta^{\un{\alpha}}$ of
the
target superspace. Other equations will put the Lagrange multipliers
$P_{\un a}{}^{ \alpha}$ to zero on shell. We shall also show that the
scalar
field $\phi$ appearing in $S_{WZ}$ is not present in $S_{GD}$, so the
derivation  of its field equation from $S_{WZ}$ in subsection
\ref{WZT}  was
correct. Thus the
superstring action \p{action} will be reduced to the Green-Schwarz
one
\p{3.26}.

In two-dimensional light-cone notation the term $S_{GD}$ becomes (see
\p{flat1form})
\bea
S_{GD} &=& \int d^2\xi D_+D_-
\left[ P_{-\un a} (D_+ X^{\un a} -i D_+
\Theta
\gamma^{\un a} \bar\Theta) +{\rm c.c.} + (+\leftrightarrow
-)\right]_{\eta=0}  \nn\\
&=& \int d^2\xi \left[D_+D_- P_{-\un a} (D_+ X^{\un a} -i D_+
\Theta\gamma^{\un a} \bar\Theta )\right.\nn\\
&& \phantom{\int d^2\xi} - D_+P_{-\un a}(D_-D_+ X^{\un a} -i D_-D_+
\Theta\gamma^{\un a} \bar\Theta  -i D_+
\Theta\gamma^{\un a} D_-\bar\Theta )\nn\\
&& \phantom{\int d^2\xi} + D_-P_{-\un a}(iD_{++} X^{\un a} +D_{++}
\Theta\gamma^{\un a} \bar\Theta -i D_+
\Theta\gamma^{\un a} D_+\bar\Theta )\nn\\
&& \phantom{\int d^2\xi} + P_{-\un a}(D_+D_-D_+ X^{\un a} -i
D_+D_-D_+
\Theta\gamma^{\un a} \bar\Theta\nn\\
&& \left.\phantom{\int d^2\xi P_{-\un a}} + D_{++}
\Theta\gamma^{\un a} D_-\bar\Theta +2i D_-D_+
\Theta\gamma^{\un a} D_+\bar\Theta) +{\rm c.c.} -
(+\leftrightarrow -)\right]_{\eta=0}
\label{4.2}
\eea
The variation with respect to the components $D_+D_- P_{-\un a}$ and
$D_+D_- P_{+\un a}$ gives two equations for the auxiliary odd
components of
the superfield $X^{\un a}$:
\be\label{4.3}
D_\pm X^{\un a}  = i D_\pm
\Theta\gamma^{\un a} \bar\Theta +i D_\pm
\bar\Theta\gamma^{\un a} \Theta
\ee
(from here on we shall drop the indication $\eta=0$). The variation
with
respect to the components $D_+ P_{-\un a}$ and $D_- P_{+\un a}$ gives
equations for the auxiliary even component $D_-D_+X^{\un a}$ and also
leads to the twistor constraint
\be\label{4.4}
D_+ \Theta\gamma^{\un a} D_-\bar\Theta +  D_+ \bar\Theta\gamma^{\un
a}
D_-\Theta = 0.
\ee
This is just the lowest-order component of the constraint \p{m3}. The
other two constraints, the $\eta=0$ components of \p{m1} and \p{m2},
follow from the terms with $D_-P_{-\un a}$ and $D_+P_{+\un a}$. This
set
of constraints was discussed in subsection \ref{WZT}. Postponing once
again the investigation of the singular solution of the constraint
till
subsection \ref{sing}, we consider only the regular one \p{matrsol}:
\be\label{4.5}
D_+\Theta^{\un\alpha} = e^{{i}\phi} \lambda_+{}^{\un\alpha}, \ \
D_-\Theta^{\un\alpha} =i e^{{i}\phi} \lambda_-{}^{\un\alpha}.
\ee

Next we shall simplify the term with $P_{-\un a}$ in \p{4.2}.
Using \p{4.3} and the anticommutation relations \p{1.1} we find that
the
first two terms after  $P_{-\un a}$ equal the third term. Further,
the
covariant derivative $D_{++}$ in this third term can be written out
in
detail according to \p{1.3}:
$$
D_{++} \Theta\gamma^{\un a} D_-\bar\Theta + D_{++}
\bar\Theta\gamma^{\un a}
D_-\Theta
$$
\be
=i e_{++}{}^m( -e^{-{i}\phi} \partial_m\Theta + e^{{i}\phi}
\partial_m\bar\Theta) \gamma^{\un
a}\lambda_-  + 2\psi_{++}{}^-\lambda_- \gamma^{\un a}\lambda_-,
\label{4.6}
\ee
where we have used \p{4.4} and \p{4.5}. Note that the covariant
derivative
$D_{++}$ in the $D_-P_{-\un a}$ term in \p{4.2} does not contain the
gravitino field, as a consequence of \p{4.3} (see also \p{pb}).

Now we are going to put all this back into the action term \p{4.2}.
The
purely auxiliary terms drop out and $S_{GD}$ is reduced to
\bea
&&S_{GD} = \int d^2\xi \left\{P_{--\un a}({\cal E}_{++}{}^{\un a}-2
\lambda_+\gamma^{\un a}\lambda_+)\right.\label{4.7}\\
&+& 2P_{-\un a}[i\chi\gamma^{\un a}\lambda_+ +
i e_{++}{}^m( -e^{-{i}\phi} \partial_m\Theta + e^{{i}\phi}
\partial_m\bar\Theta) \gamma^{\un
a}\lambda_- + 2\psi_{++}{}^-\lambda_- \gamma^{\un a}\lambda_-] -
(+\leftrightarrow -)\left.\right\},\nn
\eea
where ${\cal E}_{++}{}^{\un a}$ was defined in \p{pb} and we have
introduced the notation
\be
iD_-P_{-\un a} = P_{--\un a}, \ \ \chi = e^{-{i}\phi} D_-D_+\Theta
+e^{{i}\phi}  D_-D_+\bar\Theta
\ee
(and similarly in the $(+\leftrightarrow -)$ sector). The field
equation
for $\chi$ is a typical twistor equation,
\be\label{4.8}
P_{-\un a}(\gamma^{\un a}\lambda_+)_{\un\alpha} = 0,
\ee
which has the general solution
\be\label{4.9}
P_{-\un a} = P_{-3}(\lambda_+\gamma^{\un a}\lambda_+)
\ee
with an arbitrary odd scalar field $P_{-3}(\xi)$. Further, the
gravitino
field $\psi_{++}{}^-$ appears only in \p{4.7} (see \p{3.21}), so its
field
equation is
\be\label{4.10}
P_{-\un a}(\lambda_-\gamma^{\un a}\lambda_-) =
P_{-3}(\lambda_+\gamma_{\un
a} \lambda_+)(\lambda_-\gamma^{\un a}\lambda_-) = 0.
\ee
As explained in \p{indmet}, under the current assumption of
non-singularity of the twistor matrix the twistor factor in \p{4.10}
is
non-vanishing, so we conclude
\be\label{4.11}
P_{-3} = 0 \ \Rightarrow \ P_{-\un a} = 0.
\ee
Before inserting \p{4.11} back into \p{4.7} and thus eliminating the
$P_{-\un a}$ term from the action, we have to study the field
equation
for $P_{-\un a}$ itself:
\be\label{4.12}
i\chi\gamma^{\un a}\lambda_+ +
i e_{++}{}^m( -e^{-{i}\phi} \partial_m\Theta + e^{{i}\phi}
\partial_m\bar\Theta) \gamma^{\un
a}\lambda_- + 2\psi_{++}{}^-\lambda_- \gamma^{\un a}\lambda_- = 0.
\ee
These are three equations (as many as the projections of the vector
index
$\un a$). Two of them can be used to solve for the auxiliary field
$\chi^{\un\alpha}$ (because here we assume that the matrix
$\lambda_\pm{}^{\un\alpha}$ is
invertible). The third one enables us to solve for the
gravitino field $\psi_{++}{}^-$
(to this end one multiplies eq.\p{4.12} by $\lambda_+\gamma_{\un
a} \lambda_+$ and uses the non-singularity of the twistor factor
$(\lambda_+\gamma_{\un a} \lambda_+)(\lambda_- \gamma^{\un
a}\lambda_-)$).
Thus we see that the gravitino field is an {\it auxiliary} filed. It
is
expressed in terms of the derivative $e_{++}{}^m\partial_m\theta$
(where
$\theta = \Theta\vert_0$). This
is possible since $\theta$ transforms inhomogeneously under the
worldsheet
local supersymmetry, $\delta \theta^{\un{\alpha }} =
\epsilon^\alpha \lambda_\alpha{}^{\un{\alpha }}$ (see \p{susy'}).

So far we have shown that the term with $P_{-\un a}$ in \p{4.7} is
purely
auxiliary and drops out of the action. Now we shall show that the
term
with $P_{--\un a}$ vanishes on shell as well. First we shall vary
with
respect to the twistor field $\lambda_+$. It appears only once (we
have
already put $P_{-\un a}=0$ and in the Wess-Zumino term \p{4.7} we
have
eliminated the twistors in favor of the vectors ${\cal
E}_{\pm\pm}{}^{\un
a}$), so we get an equation similar to \p{4.8}:
\be\label{4.13}
P_{--\un a}(\gamma^{\un a}\lambda_+)_{\un\alpha} = 0 \ \Rightarrow \
P_{--\un a} = P_{-4}(\lambda_+\gamma_{\un a}\lambda_+) .
\ee
Further, the variation with respect to $P_{--\un a}$ gives
\be\label{4.14}
{\cal E}_{++}{}^{\un a} = 2 \lambda_+\gamma^{\un a}\lambda_+\;.
\ee
Finally, we vary with respect to the vielbein fields
$e_a{}^m$. They appear both in $S_{GD}$ \p{4.7} and in $S_{WZ}$
\p{3.26}.
The variational equation for $e_m{}^{--}$ is
\be\label{4.15}
P_{--\un a} {\cal E}_{m}{}^{\un a} \sim e_m{}^{++}{\cal
E}_{++}{}^{\un a}
{\cal E}_{--\un a} - {\cal E}_{m}{}^{\un a}{\cal E}_{--\un a}.
\ee
Multiplying eq.\p{4.15} by $e_{\pm\pm}{}^m$ we find
\be\label{4.16}
P_{--\un a} {\cal E}_{++}{}^{\un a} = 0, \ \ P_{--\un a} {\cal
E}_{--}{}^{\un a} \sim {\cal E}_{--}{}^{\un a}{\cal E}_{--\un a}.
\ee
Inserting the solution \p{4.13} and the $--$ analog of \p{4.14} into
\p{4.16}, we finally obtain
\be\label{4.17}
P_{-4}(\lambda_+\gamma_{\un a}\lambda_+)(\lambda_- \gamma^{\un
a}\lambda_-) = 0 \ \Rightarrow \ P_{-4} = 0 \ \Rightarrow \ P_{--\un
a} =
0.
\ee
Once again, we see that the zweibeins play the r\^ole of auxiliary
fields
(like the gravitino above). In the standard superstring theory they
produce the Virasoro constraints \p{lightl}. In the twistor theory
these
constraints are already solved in terms of twistors. Therefore the
zweibeins just give rise to auxiliary equations like \p{4.15}, which
help
eliminate some of the Lagrange multipliers.

This concludes our demonstration that the term $S_{GD}$ \p{4.1} in
the
superstring action is purely auxiliary. It does not lead to any new
equations of motion for the physical fields $x$ and $\theta$ and thus
the
on-shell component action is just the Green-Schwarz one \p{3.26}.

\subsection{The case of a degenerate twistor matrix}\label{sing}

In subsections \ref{WZT} and \ref{geo-d} we studied the component
content
of the twistor superstring action under the assumption that the
twistor
algebraic constraint \p{2.3} (taken at $\eta=0$) has the regular
solution
\p{matrsol}. Here we shall
investigate the alternative singular solution. We shall show that in
this
case the string collapses into a particle. For simplicity we shall
only
consider the bosonic fields in the action.

As explained in the Appendix, the singular solution, for which $\det
\parallel {\cal E}_\alpha{}^{\un\alpha}\parallel = 0$, has the form
\be\label{singsol}
{\cal E}_+{}^{\un\alpha} = \lambda^{\un\alpha}, \ \ {\cal
E}_-{}^{\un\alpha} = ir\lambda^{\un\alpha}.
\ee
Here $\lambda^{\un\alpha}$ is an arbitrary {\it complex} spinor and
$r$ is
an arbitrary {\it real} factor. Let us insert this solution into the
Wess-Zumino term of our string action. The quantity $A$ \p{A'}
vanishes due to the gamma matrix identity \p{gammaid}:
\be
A \sim ({\cal E}_-\gamma^{\un a}{\cal E}_+)({\cal E}_-\gamma_{\un
a}\bar{\cal
E}_+) + {\rm c.c.} = r^2 (\lambda\gamma^{\un
a}\lambda)(\lambda\gamma_{\un
a}\bar\lambda)+ {\rm c.c.} =0.
\ee
Further, the two-form term in \p{3.20} is proportional to $\theta$,
so it
does not contribute to the bosonic terms in the action. Thus,
$S_{WZ}$
vanishes in this case.

Let us now turn to the geometro-dynamical term \p{4.1}. Dropping the
fermion
fields and using the solution \p{singsol}, we see that the component
expansion in \p{4.2} is reduced to two terms only:
\be\label{volk}
S_{GD} = \int d^2\xi \left[ P_{\un a}{}^{++} (D_{++}X^{\un a}
-\lambda\gamma^{\un
a}\bar\lambda) + P_{\un a}{}^{--} (D_{--}X^{\un a}
-r^2\lambda\gamma^{\un
a}\bar\lambda) \right].
\ee
The variation with respect to the following
combination of Lagrange multipliers, $\delta P_{\un a}{}^{--} -
r^2 \delta P_{\un a}{}^{++}$ shows that
the two vectors $D_{++}X^{\un a}$ and $D_{--}X^{\un a}$
 tangent to the string surface
 are linearly dependent
\be
r^2D_{++}X^{\un a}-D_{--}X^{\un a}=0.\label{lindep}
\ee
This means that the dependence on the one of the  worldsheet
coordinates
drops out and the string collapses into a one-dimensional object
(particle).  The degeneracy of the worldsheet
leads to an additional gauge invariance. For instance, the action
\p{volk}
has the following gauge invariance
\bea\label{add}
\delta e_{++}{}^m&=&\rho(\xi)r^2e_{--}{}^m, \;\;
\delta\lambda^\alpha={1\over 2}\rho r^2\lambda^\alpha,\;\;
\delta e_{--}{}^m=\rho r^2 e_{--}{}^m, \;\;  \\
\delta P_{\un a}^{--}&=&-\rho P_{\un a}^{++} -\rho r^2 P_{\un
a}^{--}, \;\;
\delta P_{\un a}^{++}=0. \nn
\eea
 The appearence of new
gauge invariances is observed in the fermionic part of the
superstring
action too. Thus, for example, the worldsheet gravitino drops out
from the
action. This is in agreement with previous twistor formulations of
the
superparticle (see, e.g., \cite{GS}), where one does not need a
gravitino
field to achieve the local worldsheet supersymmetry invariance.

Using all these gauge invariances along with worldsheet
reparametrizations,
tangent space Lorentz and Weyl transformations, we can gauge away the
zweibeins and the field $r$. Then with the help of \p{lindep} we find
\be
S_{GD} = \int d^2\xi\; P_{\un a}(\partial_\tau X^{\un a} -
\lambda\gamma^{\un
a}\bar\lambda),
\ee
where $P_{\un a}$ corresponds to an orthogonal combination of the
Lagrange
multipliers. Integrating out the inessential worldsheet  coordinate
($\sigma$),
we see that this is a twistor particle action of the  type described
in
section \ref{part}.

The conclusion of this subsection is that when one employs the
singular
solution of the twistor constraint \p{2.3}, the superstring action
becomes degenerate. The gauge invariance widens, leaving a number of
component fields arbitrary. The remaining physical fields do not
depend on
$\sigma$ any more, so the superstring becomes a superparticle. Since
the
ordinary Green-Schwarz superstring formulation does contain the
superparticle as a certain singular limit,
we see that both the regular and singular solutions to the
twistor constraints have to be taken into account.

\section{Conclusions}

In this paper we have shown how the non-heterotic $D=3$ type II
superstring can be formulated with manifest $N=(1,1)$ worldsheet
supersymmetry. The central point in the construction was the
geometro-dynamical constraint \p{2.1} and its corollary \p{2.3}. In
particular, they reduced the initial $2\times 2$ complex twistor
matrix
${\cal E}_\alpha{}^{\un\alpha}$ to the two null vectors from the
Virasoro
constraints. The rest of \p{2.1} gave rise to purely auxiliary
equations.

The geometro-dynamical principle is common for the twistor
formulations of
the superparticle \cite{GS}, the heterotic superstring \cite{TON},
\cite{DGHS} and, as
we have seen here, the non-heterotic $D=3$ type II superstring. One
would
be tempted to extrapolate this to the non-heterotic type II
superstring in
higher dimensions as well. Indeed, analysing the lowest-order
component of
eq. \p{2.3}, one can show that the $D=3$ situation is reproduced. For
instance, in $D=10$ the $16\times 16$ complex twistor matrix is once
again reduced to the two null vectors from the Virasoro constraints.
However, starting from $D=4$ (and $N=(2,2)$)
there is an unexpected difficulty at the next level in the $\eta$
expansion of eq. \p{2.3}. One can show (most easily in the linearized
approximation) that some of the constraints are equations of motion
for
$\theta$.  This is inadmissible, since the geometro-dynamical
constraint
is produced by a Lagrange multiplier, which implies that some of the
components of the latter will propagate as well.  One clearly sees
that the
case $D=3$ is the only exception, due to the trivial algebra of the
transverse gamma matrices in $D=3$. In fact, the same problem is also
encountered in the framework of the type II superparticle discussed
in
section \ref{part}.  So, the main open problem now is to find a
modification of the geometro-dynamical constraint such that it would
not
imply equations of motion in $D>3$. We hope to be able to report
progress
in this direction elsewhere.

\vskip15mm {\bf Note added} After this paper has been completed, we
received a new preprint by Pasti and Tonin \cite{PTo}, in which they
claim
that a similar construction applies to the $D=11$ supermembrane with
full
$N=8$ $D=3$ worldsheet supersymmetry. This would be very surprising,
since
they impose the same type of geometro-dynamical constraint. As we
mentioned
above, in the case of extended ($N>1$) worldsheet supersymmetry this
constraint is most likely to produce equations of motion
 and the corresponding Lagrange multiplier will contain new
 propagating
degrees of freedom. One
simple argument explaining this phenomenon has been proposed to us by
P. Howe.
The supermembrane theory of \cite{PTo} could be truncated to a $D=11$
superparticle with $N=16$ worldline supersymmetry. There the
geometro-dynamical constraint reduces the twistor variables (i.e. the
bosonic physical fields) to the sphere $S^9$ (modulo gauge
transformations). At the same time, the 32 components of the fermion
$\theta^{\un\alpha}$ are brought down to 16 after taking into account
the
16 local worldline supersymmetries. It is clear that 9 bosons and 16
fermions do not form an off-shell supermultiplet, therefore the
geometro-dynamical constraint must involve equations of motion.

\section{Appendix.  Solution to the twistor constraints}

In section \ref{dyn} we derived the geometro-dynamical constraint
\p{2.3} or, in light-cone notation, \p{m1}-\p{m3}. The lowest-order
terms
in the $\eta$ expansion of this constraint gives restrictions on the
twistor matrix $\parallel {\cal E}_\alpha{}^{\underline\alpha}
\parallel$:
\bea\label{m11}
&&{\cal E}_+\gamma^{\un a} \bar {\cal E}_+ = {\cal E}_{++}{}^{\un a},
\\
&&{\cal E}_-\gamma^{\un a} \bar {\cal E}_- = {\cal E}_{--}{}^{\un a},
\label{m12}\\
&&{\cal E}_+\gamma^{\un a} \bar {\cal E}_- + {\cal E}_-\gamma^{\un a}
\bar
{\cal E}_+ = 0. \label{m13}
\eea
In fact, the first two equations define two vectors ${\cal
E}_{\pm\pm}{}^{\un a}$ and only the third equation constrains the
twistor
variables. Here we are going to solve \p{m13} in a general way.

We start by writing out the components of the twistor matrix
\be\label{A1}
{\cal E}_\alpha{}^{\underline\alpha} \equiv
\left(\matrix{A & B\cr C & D \cr}\right).
\ee
A basic kinematic assumption about the twistor variables is that they
can
never vanish identically. This means that at least one element of the
matrix \p{A1} is non-vanishing.
It is convenient to write down the constraint \p{m13} using the
light-cone basis \p{A3} for the gamma matrices.
There the three projections read
\bea
(++):&&  A\bar C + C\bar A = 0 , \label{con1} \\
(--):&&  B\bar D + D\bar B = 0 ,\label{con2}\\
(+-): &&  B\bar C +C\bar B + A\bar D + D\bar A = 0 \ . \label{con3}
\eea
The general solution to equations \p{con1}, \p{con2} is given by
\be
A=ae^{i\alpha}, \;\; C=ice^{i\alpha}, \;\;
B=be^{i\beta}, \;\; D=id e^{i\beta},
\label{gen}\ee
where $a,b,c$ and $d$ are {\it real}.  Substituting this into
\p{con3} one gets
\be\label{last}
(ad-bc)\sin(\alpha-\beta)=0.
\ee
Now, there are two possibilities: the matrix
${\cal E}_\alpha{}^{\underline\alpha}$ can be either degenerate or
non-degenerate. With the help of \p{gen} we evaluate the determinant
of
this matrix
\be
\det \parallel{\cal E}_\alpha{}^{\underline\alpha}\parallel=
i(ad-bc)e^{i(\alpha+\beta)}.
\label{det}\ee

Hence if the matrix ${\cal E}_\alpha{}^{\underline\alpha}$ is
non-degenerate,
$ad-bc\neq 0$ and \p{last} implies in turn $\alpha=\beta$. In the
degenerate case $ad-bc=0$ (and hence
$(c,d) \sim (a,b)$) and the phases $\alpha$ and $\beta$ are
independent.

In summary, the general solution to \p{m13} consists of two sectors.
In the
first sector, the matrix ${\cal E}_\alpha{}^{\underline\alpha}$ is
{\it
non-degenerate} and represented as follows
\be
{\cal E}_\alpha{}^{\underline\alpha}=e^{i\phi}
\left(\matrix{a & b\cr ic & id \cr}\right)\equiv
e^{i\phi} \left(\matrix{\lambda_+^\alpha\cr i\lambda_-^\alpha
\cr}\right),
\label{nondeg}\ee
where the spinors $\lambda_+^\alpha$ and $\lambda_-^\alpha$ are {\it
real}
and restricted by the condition
\be\label{rest}
\lambda_+^\alpha\lambda_{-\alpha}\equiv ad-bc\neq 0.
\ee

The second sector consists of the {\it degenerate} matrix
${\cal E}_\alpha{}^{\underline\alpha}$
\be
{\cal E}_\alpha{}^{\underline\alpha}=
\left(\matrix{ae^{i\alpha} & be^{i\beta}\cr
irae^{i\alpha} & irbe^{i\beta} \cr}\right)\equiv
\left(\matrix{ \lambda^\alpha\cr ir\lambda^\alpha \cr}\right)
\label{degen}\ee
where $\lambda^\alpha$ is now an arbitrary {\it complex} spinor,
and $r$ is {\it real }.

\vskip15mm
{\bf Acknowledgements}

E.S. is grateful to N. Berkovits, F. Delduc, P. Howe, E. Ivanov, W.
Nahm,
A. Perelomov, M. Scheunert for stimulating discussions.

\end{document}